\documentclass[twocolumn,nofootinbib,amsmath]{revtex4}
 \usepackage{color}
\begin{document}

\title{Rigid particle revisited: extrinsic curvature yields the Dirac equation}

\author{Alexei Deriglazov} \email{alexei.deriglazov@ufjf.edu.br}

\affiliation{Depto. de Matem\'atica, ICE, Universidade Federal de Juiz de Fora, MG, Brazil}

\author{Armen Nersessian} \email{arnerses@ysu.am}
\affiliation{Yerevan State University,
1 Alex Manoogian St., Yerevan, 0025, Armenia}

\begin{abstract}
We reexamine the model of relativistic particle with higher-derivative term linear on the first extrinsic curvature
(rigidity). The passage from classical to quantum theory requires a number of rather unexpected steps which we report
here. We found that, contrary to common opinion, quantization of the model in terms of $so(3.2)$-algebra yields massive
Dirac equation.
\end{abstract}

\maketitle 
\section{Introduction}
At the end of eighties much attention has been paid to the study of relativistic mechanical models with Lagrangians
depending on extrinsic curvatures of the world line. These studies were mostly inspired by the Polyakov's papers on
rigid strings \cite{PolyakovRig} and Chern-Simons theories \cite{PolyakovCS}. Initially these models were considered as
toy models for the above mentioned field-theoretical ones, but very soon it was realized that they are of their own
interest, and probably could be considered as mechanical models of relativistic spin (see, e.g.
\cite{86}-\cite{hindu}). The first system of this sort has been suggested by Pisarski \cite{86} as a toy model of rigid
string. It is given by the action
\begin{eqnarray}\label{2.1}
S=\int d\tau\sqrt{-\dot x^2}\left[-mc+c_1k_1(\dot x, \ddot x)\right],
\end{eqnarray}
where $k_1$ is  the first  extrinsic  curvature of world line
\begin{eqnarray}\label{0.1}
k_1=\frac{\sqrt{-(\dot x^2\ddot x^2-(\dot x\ddot x)^2)}}{|\dot x|^3}, \quad |\dot x|=\sqrt{-\dot x^2}.
\end{eqnarray}
This system,  as well as other three- and four-dimensional systems with the Lagrangians  depending on extrinsic
curvatures were investigated by many authors. It has been observed that when $m=0$, it describes massless particle with
the helicity $c_1$\cite{mishamassless}, while the case $m\ne 0$, $c_1\ne 0$ implies a model with ten-dimensional phase space, which on the
quantum level does not describe an elementary system (that is an irreducible representation of Poincar\'e group)\cite{misharigid}.
Further studies of similar systems in arbitrary space-time \cite{ner1}, as well as on null-curves \cite{null} result in
the same conclusion: only the actions proportional to single extrinsic curvature yield, upon quantization, the massless
irreducible representations of Poincar\'e group. It was also established that the mass term in (\ref{2.1}) prohibits
the constraint which could be classical analog of Klein-Gordon equation.

However, Klein-Gordon equation could follows from the Dirac one at the {\sl quantum} level, so that it is not necessary
to have an analog of the later equation at the classical level \cite{AAD3, AAD4}.
Besides, the conclusion on reducibility of quantum space of states has been made from analysis of constraints of
classical theory. Here we use an ambiguity in the passage from classical to quantum theory which was unnoticed in
previous works. This allows us to fix the second Casimir operator (\ref{ad3}) of Poincare group.

In the present work, following this ideology, we analyze the action (\ref{2.1}) with $m\ne 0$ and
$c_1=\frac{\sqrt{3}\hbar}{2}$, and show that its canonical quantization leads to the Dirac equation
with the mass
\begin{eqnarray}\label{2.2}
M=\frac{\sqrt{3}}{2}m.
\end{eqnarray}
Any other choice of $c_1$ turns out to be inconsistent with our quantization procedure, see below.
So, in contrast with common opinion, quantization of rigid particle given by the action (\ref{2.1}) results in the
elementary system of spin one-half.

\section{Hamiltonian formulation}
Consider the time-like curve $x^\mu(\tau)$ (parameterized by arbitrary $\tau$) in four-dimensional Minkowski space
$\eta^{\mu\nu}=(-, +, +, +)$: ${\dot x}^2<0$.

Let us consider the action \eqref{2.1}, denoting, for convenience, $-mc=c_0$. To be able to construct Hamiltonian
formulation of the theory, we first use the Ostrogradsky method and represent our higher-derivative Lagrangian as a
Lagrangian
 with first-order derivatives. That is we represent (\ref{2.1}) in the form
\begin{eqnarray}\label{2.3}
S=\int d\tau\left[ c_0|\omega|+c_1\frac{\sqrt{\dot\omega N(\omega)\dot\omega}}{|\omega|}+ p(\dot x-\omega)\right]\; ,
\end{eqnarray}
where we  have introduced the projector
\begin{eqnarray}\label{0.2}
N^{\mu\nu}=\eta^{\mu\nu}-\frac{\omega^\mu\omega^\nu}{\omega^2}: \quad \omega_\mu N^{\mu\nu}=0, \qquad N^2=N.
\end{eqnarray}
Let us construct Hamiltonian formulation of the model. We denote the conjugated momenta for $\omega^\mu$ as $\pi_\mu$,
and the conjugated momenta for $x^\mu$ by $p_\mu$. Applying the standard machinery \cite{GT1, AAD8} we get that the
system possesses two primary constraints
\begin{eqnarray}\label{2.5}
\omega\pi=0, \qquad \omega^2\pi^2+c_1^2=0,
\end{eqnarray}
and the Hamiltonian
\begin{eqnarray}\label{2.7}
H=p\omega-c_0|\omega|+v_1(\omega^2\pi^2+c_1^2)+v_2\omega\pi,
\end{eqnarray}
where $v_i$ are Lagrangian multipliers associated with the primary constraints. Note, that  Eq. \eqref{2.5} together
with $\omega^2<0$ imply $\pi^2>0$.

Combining the constraints, we could replace $\omega^2\pi^2=-c_1^2$ by the equivalent constraint
$J^{\mu\nu}J_{\mu\nu}=-8c_1^2$, where $J^{\mu\nu}=2\omega^{[\mu}\pi^{\nu]}$ is angular momentum of $(\omega,
\pi$)\,-space. As each motion take place with the same angular momentum, we expect that quantum theory describes an
elementary system of fixed spin.

Computing derivatives of the primary constraints, and so on, we get the following sets of the first-class constraints
\begin{eqnarray}\label{2.9}
\omega\pi=0, \quad \Rightarrow \quad p\omega-c_0|\omega|=0,
\end{eqnarray}
the second-class ones
\begin{eqnarray}\label{2.8}
\omega^2\pi^2+c_1^2=0, \quad \Rightarrow \quad p\pi=0,
\end{eqnarray}
as well as the equation determining the Lagrangian multiplier,
$v_1=-|\omega|\frac{p^2+c_0^2}{2c_0c_1^2} \label{v1}$.
The multiplier $v_2$ remains an arbitrary function, in accordance with reparametrization invariance of the action
(\ref{2.1}) \cite{hindu}.

To take into account the second-class constraints (\ref{2.8}), we pass from Poisson to Dirac bracket. Nonvanishing
Dirac brackets are
\begin{eqnarray}\label{2.12}
\{x^\mu, p^\nu\}=\eta^{\mu\nu}, \quad \{\omega^\mu, \omega^\nu\}=\frac{\omega^2}{\pi^2(p\omega)}p^{[\mu}\pi^{\nu]},\cr
\{\omega^\mu, \pi^\nu\}=\eta^{\mu\nu}-\frac{p^\mu\omega^\nu}{p\omega}, \qquad \{\pi^\mu, \pi^\nu\}=0, \cr \{\omega^\mu,
x^\nu\}=-\frac{\omega^2}{\pi^2(p\omega)}\pi^\mu\pi^\nu, \quad \{\pi^\mu, x^\nu\}=\frac{1}{(p\omega)}\omega^\mu\pi^\nu.
\end{eqnarray}
So, we are ready to quantize our model. We could quantize the Dirac brackets (\ref{2.12}) and impose the first-class
constraints (\ref{2.9}) as operator equations on quantum states, thus obtaining some equations for the wave function.

\section{Quantization}
The function $v_2$ was not determined through the Dirac procedure, and enters as arbitrary function into general
solution to equations of motion \cite{nesterenko, pav} for $x$ and $\omega$. Dynamics of basic variables is ambiguous,
which is reflected by invariance of the action (\ref{2.3}) under local transformations studied in details in
\cite{tmp}. So we pass from initial gauge non-invariant variables to the set of candidates for obsevables. Then we show
that quantization of the set admits a reasonable interpretation of resulting quantum theory as a model of spin one-half
particle.

Namely, we note that in the model there is the set of phase-space functions
\begin{eqnarray}\label{2.13}
\tilde J^{5\mu }\equiv 2c_1K^\mu_\alpha\left(\frac{\omega^\alpha}{|\omega|}+\frac{p^\alpha}{c_0}\right),
\end{eqnarray}
\begin{eqnarray}\label{2.14}
\tilde J^{\mu\nu}\equiv 2K^{\mu}_\alpha\left(\omega^\alpha+\frac{|\omega|p^\alpha}{c_0}\right)\pi^\beta K^\nu_\beta-
(\mu\leftrightarrow\nu),
\end{eqnarray}
where
\begin{eqnarray}\label{2.18}
K^{\mu}_{\alpha}=\delta^{\mu}_\alpha- Ap^\mu p_\alpha, \qquad A=\frac{\sqrt{p^2+c_0^2}-|c_0|}{p^2\sqrt{p^2+c_0^2}}.
\end{eqnarray}
Their  Dirac brackets generate, on the first-class constraint surface, the  $so(3.2)$\,-algebra
\begin{equation}\label{2.16}
\begin{array}{c}
\{\tilde J^{5\mu }, \tilde J^{5\nu }\}=-2\tilde J^{\mu\nu}, \\
 \{\tilde J^{\mu\nu}, \tilde J^{5\alpha }\}=2\eta^{\mu\alpha}J^{5\nu }-2\eta^{\nu\alpha}J^{5\mu}, \\
 \{\tilde J^{\mu\nu}, \tilde J^{\alpha\beta}\}=2(\eta^{\mu\alpha}\tilde J^{\nu\beta}-
\eta^{\mu\beta}\tilde J^{\nu\alpha}- \eta^{\nu\alpha}\tilde J^{\mu\beta}+\eta^{\nu\beta}\tilde J^{\mu\alpha}).
\end{array}
\end{equation}
To see that this is indeed  $so(3.2)$, let us denote $\tilde J^{AB}=(\tilde J^{5\mu}, \tilde J^{\mu\nu})$ and introduce
the five-dimensional metric $\eta^{AB}=(-,+,+,+,-)$.
 Then the algebra (\ref{2.16}) acquires the form
\begin{equation}\label{2.17}
\begin{array}{c}
\{ \tilde J^{AB},  \tilde J^{CD}\}=\\
2(\eta^{AC} \tilde J^{BD}-\eta^{AD} \tilde J^{BC}-  \eta^{BC} \tilde J^{AD}+\eta^{BD} \tilde J^{AC}),
\end{array}
\end{equation}
which is just $so(3.2)$\,-algebra.

The first-class constraints (\ref{2.9}) can be presented through the new variables as follows
\begin{equation}\label{ad1}
p_\mu \tilde J^{5\mu}+2c_1\sqrt{p^2+c_0^2}=0,
\end{equation}
\begin{equation}\label{ad2}
(\omega\pi)^2=-c_1^2+\frac{1}{p^2}\left(\frac{1}{4}S^\mu S_\mu-c_1^2c_0^2\right),
\end{equation}
where $S^\mu=\frac12\epsilon^{\mu\nu\alpha\beta}p_\nu\tilde J_{\alpha\beta}$ is the Pauli-Lubanski vector (then $S^2$
represents Casimir operator of Poincare group).

The initial coordinates $x^\mu$ have non-zero Dirac brackets with $\tilde J^{AB}$. To improve this, we introduce the
effective coordinates $X^\mu=x^\mu+|w|\pi^\mu/c_0$ commuting with $\tilde J^{AB}$, and take operators associated with
$X^\mu$, $p_\mu$ in the standard form,
$\hat X^\mu=X^\mu$, $\hat p_\mu=-i\hbar\frac{\partial}{\partial X^\mu}$.

To quantize the spin variables, we look for operators with commutators obeying the Dirac-brackets algebra on the
constraint surface, the equation (\ref{2.16}). That is we adopt the rule
$[ ~ , ~ ]=i\hbar\left.\left.\{ ~, ~\}_{DB}\right|_{1CC}\right|_{q\rightarrow \hat q}$.
This guarantees the correspondence principle: the operators (in the Heisenberg picture) and corresponding classical
variables will obey the same equations of motion.

Choice of spin-sector operators is dictated by the constraint $\omega\pi=0$ as follows. According to Eq. (\ref{2.12}),
this contains product of variables with non-vanishing Dirac brackets, so the corresponding quantum operator will
contain product of non-commuting operators. Any two (Lorentz-invariant) operators which we can associate with the
constraint differ on a number. So, there is an ambiguity in the passage from classical to quantum theory
\begin{equation}
\omega\pi=0 ~\rightarrow ~ \hat\omega\hat\pi=c_2. \label{2.28}
\end{equation}
We propose to fix the ordering constant to be $c_2^2=-c_1^2$. Then quantum counterpart of equation (\ref{ad2}) states
that Casimir operator of the Poincare group has fixed value
\begin{equation}\label{ad3}
S^2=4c_1^2c_0^2=3\hbar^2m^2c^2,
\end{equation}
corresponding to spin one-half irreducible representation (Note that our state-space is picked out by unique equation
(\ref{ad1}). The only irreducible representation of Poincar\'e group of such a kind is the spin one-half
representation. So, any choice of $c_1$ different from $c_1=\frac{\sqrt{3}\hbar}{2}$ would lead to inconsistent
picture. Besides, similarly to string theory, other choice of $c_2$ would lead to appearance of negative norm states in
the spectrum).

Hence we are forced to quantize the variables $\tilde J^{5\mu}$ and ${\tilde J}^{\mu\nu}$ by $\gamma$-matrices,
$\tilde J^{5\mu} \rightarrow \hbar \gamma^\mu$, $\tilde J^{\mu\nu} \rightarrow \hbar \gamma^{\mu\nu}$,
where
\begin{eqnarray}\label{2.19}
\gamma^0= \left(
\begin{array}{cc}
1& 0\\
0& -1
\end{array}
\right), \quad \gamma^i= \left(
\begin{array}{cc}
0& \sigma^i\\
-\sigma^i& 0
\end{array}
\right),
\end{eqnarray}
and
\begin{eqnarray}\label{2.20}
\gamma^{\mu\nu}\equiv\frac{i}{2}(\gamma^\mu\gamma^\nu-\gamma^\nu\gamma^\mu).
\end{eqnarray}
They form a representation of $so(3.2)$
\begin{eqnarray}\label{2.21}
[\gamma^\mu, \gamma^\nu]=-2i\gamma^{\mu\nu}, \quad [\gamma^{\mu\nu},
\gamma^\alpha]=2i(\eta^{\mu\alpha}\gamma^\nu-\eta^{\nu\alpha}\gamma^\mu), ~ \cr [\gamma^{\mu\nu},
\gamma^{\alpha\beta}]= 2i(\eta^{\mu\alpha}\gamma^{\nu\beta}- \eta^{\mu\beta}\gamma^{\nu\alpha}-
\eta^{\nu\alpha}\gamma^{\mu\beta}+\eta^{\nu\beta}\gamma^{\mu\alpha}).
\end{eqnarray}
The operators act on space of Dirac spinors $\Psi_a$, $a=1, 2, 3, 4$.

Let us see the meaning of the first-class constraint (\ref{ad1}). Its quantum counterpart reads
\begin{eqnarray}\label{2.24}
\left(\gamma^\mu\hat p_\mu+\frac{2c_1}{\hbar}\sqrt{\hat p^2+c_0^2} ~\right)\Psi=0.
\end{eqnarray}
Then we obtain, as a consequence, the following Klein-Gordon equation
\begin{equation}\label{2.25}
\left(\hat p^2+\frac{4c_0^2c_1^2}{\hbar^2+4c_1^2}\right)\Psi=0.
\end{equation}
Substitute (\ref{2.25}) into (\ref{2.24}), this gives the Dirac equation $(\gamma^\mu\hat p_\mu
-2c_0c_1(\hbar^2+4c_1^2)^{-\frac{1}{2}})\Psi=0$. Taking $c_0=-mc$ and $c_1=\frac{\sqrt{3}\hbar}{2}$, we arrive at the
final result
\begin{equation}\label{2.26}
\left(\gamma^\mu\hat p_\mu +\frac{\sqrt{3}}{2}mc\right)\Psi=0.
\end{equation}
In resume, canonical quantization of the action (\ref{2.1}) in properly chosen variables (\ref{2.13}), (\ref{2.14})
leads to the theory of spin one-half particle. This observation contradicts with common opinion about role of
higher-derivative models in the description of massive spinning particles. Many questions arise in this respect. Is it
possible to introduce reasonable interaction \cite{ner1, ner2} with electromagnetic and curved backgrounds? Could other
models with Lagrangians depending on extrinsic curvatures (say, on torsion) describe massive and massless spinning
particles on  space of non-constant curvature? If so, of which kind, and of which spin? Could this revision changes our
opinion on the role of such systems in quantum optics \cite{opt1} and polymer physics \cite{bio1, bio2}? Clarification
of this questions should be subject of separate investigation.

\acknowledgments We thank A. Pupasov - Maximov for useful conversations and interest in the work. This work was
initiated during numerous discussions at Universidade Federal de Juiz de Fora, and has been mostly  done during the
stay of A.N. there. He thanks administration of the University for invitation, financial support and kind hospitality.
The work was partially supported by Brazilian agencies FAPEMIG and CNPq (A.D.), by the grant of Armenian State
Committee of Science 11-1c258, by Volkswagenstiftung under contract nr. 86 260 (A.N).

\end{document}